\documentclass[prl,reprint,floatfix,noeprint]{revtex4-1}
\usepackage{amsmath,amssymb,graphicx,braket,hyperref}
\hypersetup{colorlinks=true,urlcolor=[rgb]{0.0,0.0,0.55},citecolor=,linkcolor=}
\allowdisplaybreaks
\newcommand{\diff}[1]{\text{d}#1\,}
\newcommand{\be}{\begin{equation}}
\newcommand{\ee}{\end{equation}}
\newcommand{\bea}{\begin{eqnarray}}
\newcommand{\eea}{\end{eqnarray}}
\newcommand{\ba}{\begin{eqnarray*}}
\newcommand{\ea}{\end{eqnarray*}}
\newcommand{\dagga}{{\phantom{\dagger}}}
\newcommand{\up}{\uparrow}
\newcommand{\down}{\downarrow}

\newcommand{\Tr}{\text{Tr}}
\newcommand{\m}[1]{\mathcal{#1}}

\usepackage{xr}
\begin{document}

\title{Resonant thermalization of periodically driven
strongly correlated electrons}

\author{Francesco Peronaci}
\author{Marco Schir\'o}
\author{Olivier Parcollet}
\affiliation{Institut de Physique Th\'{e}orique (IPhT),
CEA, CNRS, UMR 3681, 91191 Gif-sur-Yvette, France}

\date{\today}

\begin{abstract}
We study the dynamics of the Fermi-Hubbard model
driven by a time-periodic modulation
of the interaction within nonequilibrium Dynamical Mean-Field Theory.
For moderate interaction, we find clear evidence of thermalization
to a genuine infinite-temperature state with no residual oscillations.
Quite differently, in the strongly correlated regime, we find 
a quasi-stationary extremely long-lived state with oscillations 
synchronized with the drive (Floquet prethermalization).
Remarkably, the nature of this state dramatically changes upon tuning the
drive frequency.
In particular, we show the existence of a critical frequency
at which the system rapidly thermalizes despite the large interaction.
We characterize this resonant thermalization and provide an analytical
understanding in terms of a break down of the
periodic Schrieffer-Wolff transformation.
\end{abstract}

\maketitle

Recent advances in the ability to tailor and control light-matter interaction
on the ultra-fast time
scale~\cite{Orenstein2012,Nicoletti2016,Forst2011,Subedi2014}
have brought increasing interest in the manipulation of quantum phases
of matter with periodic \emph{driving} fields.
Notable achievements are light-induced
superconductivity~\cite{Fausti2011,Mitrano2016},
metal-to-insulator transition~\cite{Lantz2017}
and control of microscopic parameters
such as the local interaction in organic Mott
insulators~\cite{Singla2015}
and the band gap in excitonic insulators~\cite{Mor2017}.
Similar ideas are applied to ultra-cold atoms in optical
lattices~\cite{Bloch2008,*Bloch2012} where driving fields are used,
for instance, to engineer topological states~\cite{Eckardt2017,*Jotzu2014}.

From a theoretical perspective,
periodically driven, or \emph{Floquet}, quantum systems are a
long-standing subject of
studies ranging from
dynamical localization~\cite{Dunlap1986} and
quantum dissipation~\cite{Grifoni1998} to quantum chaos~\cite{Casati2006}
and, more recently, isolated quantum many-body systems~\cite{Moessner2017}.
Other topics of active research include
drive-induced topological states~\cite{Oka2009,Lindner2011}
and artificial gauge fields~\cite{Goldman2014};
driven electron-phonon coupling~\cite{Knap2016,Babadi2017,Murakami2017};
and
integrable systems~\cite{Lange2017},
correlated electrons~\cite{Tsuji2008,*Tsuji2009,Tsuji2011,Coulthard2017a,Mazza2017}
or topological
systems~\cite{Dehghani2014,*Dehghani2015a,Dehghani2015,*Dehghani2016}
in presence of dissipation.

In absence of integrability and of many-body
localization, isolated out-of-equilibrium
quantum many-body systems are expected
to show thermalization of local observables at long times~\cite{Rigol2008}.
Driven systems, which lack time translational invariance, are therefore
brought to \emph{thermalize} to a featureless infinite-temperature state
consistent with maximum entropy and no energy
conservation~\cite{DAlessio2014,Lazarides2014,Ponte2015,Genske2015}.
Yet, the transient dynamics may leave space to non-trivial extremely
long-lived non-thermal states characterized by oscillations synchronized
with the drive, a phenomenon known as \emph{Floquet prethermalization}.
This prethermal behavior can emerge in the high frequency
limit~\cite{Bukov2015a,Abanin2015,Kuwahara2016,Bukov2016a,Mori2016,Abanin2017a,Machado2017}
or be the consequence of a nearby integrable point
in the system parameter space.
In this case, 
as recently observed for weakly~\cite{Bukov2015,Canovi2016,Weidinger2017}
 and strongly~\cite{Seetharam2017,Herrmann2017} 
interacting systems,
there are many quasi-integrals of motion
that prevent thermalization except at very long times, 
similarly to what happens after a quantum
quench~\cite{Moeckel2008}.
However, many intriguing questions remain wide open especially concerning
the intermediate coupling and frequency regimes,
where the most remarkable phenomena are expected to occur.

In this Letter we consider the Fermi-Hubbard model
as paradigmatic example of strongly correlated electrons.
The system is subject to a time-periodic modulation of
the electron interaction, but it is otherwise isolated from any external
reservoir.
Starting from a thermal equilibrium state,
we use non-equilibrium Dynamical Mean-Field Theory
(DMFT)~\cite{Aoki2014}
to calculate the time evolution induced by the drive.
First, we explicitly show that at moderate interaction the system
thermalizes to the infinite-temperature state.
Then, we turn to the regime of large interaction and find
a long-lived prethermal state synchronized
with the drive, except for a critical, resonant frequency where
we find thermalization
and a behavior reminiscent of a dynamical
transition~\cite{Eckstein2009b,Schiro2010,Tsuji2013a}.
A periodic Schrieffer-Wolff transformation
shows that the Floquet prethermalization is due to the 
quasi-conservation of double occupancy at large interaction,
with the resonant thermalization emerging in correspondence
of a break down of such an expansion.

The system is governed by the following Hamiltonian:
\begin{equation}
\label{eqn:hubb}
H(t)= \sum_{i,j}\sum_{\sigma=\up,\down}V_{ij}c^\dagger_{i\sigma}c_{j\sigma}
+U(t)\sum_i(n_{i\up}-\frac{1}{2})(n_{i\down}-\frac{1}{2}),
\end{equation}
where $U(t)=U_0+\delta U\sin\Omega t$ is the periodically driven interaction
and $V_{ij}$ is the hopping, which is such that the bare density
of states reads
$\rho(\epsilon)=\sqrt{4V^2-\epsilon^2}/(2\pi V^2)$
(Bethe lattice).
We take $V$ as unit of energy, frequency and inverse of time ($\hbar=1$).
In these units the bare band-width is $W=4$
and the critical point of the Mott transition in DMFT is at $U_c\simeq 4.8$ and
 at an inverse temperature $\beta_c\simeq20$.
We consider a thermal initial density matrix
$\rho(0)=\exp(-\beta H(0))$ with $\beta=5$
and we fix the drive amplitude $\delta U=2$ (cf. Supp.
Mat.~\footnote{See Supplemental
Materials at \dots for additional data with different interaction and
drive amplitude,
for details on the non-crossing and one-crossing approximations,
the spectral analysis
and the Schrieffer-Wolff transformation.} Sec.~\ref{amplitude}).
For all times the interaction remains repulsive and
the system stays half-filled ($\braket{n_\sigma}=0.5$) and particle-hole
symmetric.

To calculate the time evolution induced by the drive
we use nonequilibrium DMFT~\cite{Aoki2014}, which consists in 
mapping the lattice model described by Eq.~\eqref{eqn:hubb}
onto a quantum impurity problem with the following action:
\begin{equation}
\label{eqn:Seff}
\m{S}=\m{S}_\text{loc}+\int_\m{C}\diff{t}\diff{t'}
\sum_{\sigma=\up,\down}c^\dagger_{\sigma}(t)\Delta_\sigma(t,t')c_\sigma(t'),
\end{equation}
where $\m{S}_\text{loc}$ is the action associated to the local term
in Eq.~(\ref{eqn:hubb}),
$\m{C}$ is the three branch Keldysh contour~\cite{Stefanucci2013}
and $\Delta_{\sigma}(t,t')=V^2G_{\sigma}(t,t')$ is the hybridization between
the impurity and a nonequilibrium bath, which is self-consistently determined
from the impurity Green function
$G_{\sigma}(t,t')=-i\braket{T_\m{C}c_{\sigma}(t)c^{\dagger}_{\sigma}(t')}$.
Within the DMFT mapping, 
the impurity Green function coincides with the local
lattice Green function and from it we can calculate various quantities
directly in the thermodynamic limit, 
such as the double occupancy $d(t)=\braket{n_{i\up}(t)n_{i\down}(t)}$
and the kinetic energy $K(t)=\sum_{ij\sigma}V_{ij}
\braket{c^\dagger_{i\sigma}(t)c^\dagga_{j\sigma}(t)}$.
The computation of the impurity Green function
is a challenging task and,
despite recent progresses~\cite{Cohen2015,Antipov2017,Profumo2015},
 an efficient and numerically exact approach is still lacking.
Here we resort to the non-crossing
approximation~\cite{Bickers1987,Nordlander1999,Ruegg2013,Eckstein2010,
Eckstein2010a,Eckstein2011a,*Eckstein2013,Strand2015a,Golez2015}
which consists in a first order self-consistent
hybridization expansion and which we implement through a Dyson equation
for the impurity atomic-state propagator
(cf. Supp. Mat.~\cite{Note1} Sec.~\ref{nca_oca}).
For moderate interaction, we benchmark the results with the next-order
one-crossing approximation
(cf. Supp. Mat.~\cite{Note1} Sec.~\ref{benchmark}).

\begin{figure}
\includegraphics[width=\columnwidth]{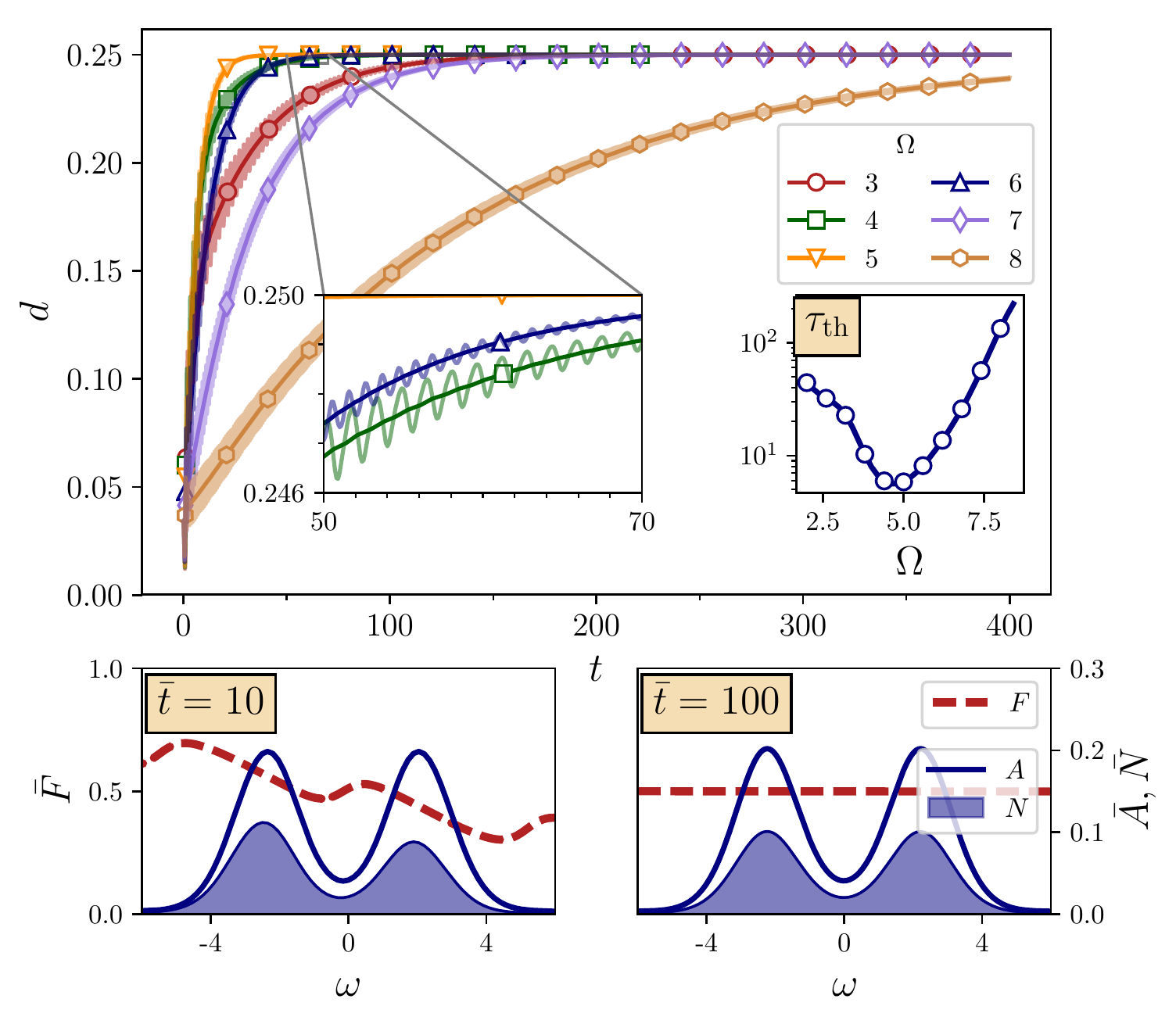}
\caption{\label{fig:fig1}Thermalization to infinite temperature
($U_0=4$).
Top panel: Double occupancy $d(t)$ for various drive frequencies~$\Omega$
shows oscillations (shade)
on top of an exponential relaxation (solid line).
Right inset: Thermalization time $\tau_\text{th}(\Omega)$.
Bottom panels: Averaged spectral function $\bar A(\omega,\bar t)$,
occupation function $\bar N(\omega,\bar t)$ and
distribution function
$\bar F(\omega,\bar t)=\bar N(\omega,\bar t)/\bar A(\omega,\bar t)$
for~$\Omega=4$ show the evolution from the out-of-equilibrium
state at~$\bar t=10$ to the infinite-temperature thermal state at~$\bar t=100$.}
\end{figure}

We start by discussing the results for moderate average
interaction $U_0=4$ (Fig.~\ref{fig:fig1}).
The double occupancy shows fast oscillations with frequency comparable to the
one of the drive~$\Omega$ superimposed to a slower but exponential relaxation.
Quite interestingly, after the initial transient and despite the continuous
driving, the oscillations get fully damped and the double occupancy
reaches the value~$d_\text{th}=0.25$
independently of the frequency.
This is the value of a maximally disordered
state and as such signals the thermalization to infinite temperature.
With an exponential fit we can extract the thermalization
time~$\tau_\text{th}$
which is minimum for $\Omega\simeq 4.8$ and diverges
for large frequency.

Thermalization is confirmed by the evolution of the Green function and
in particular of the retarded component $G^{R}_{\sigma}(t,t')=-i\theta(t-t')
\braket{\left\{c_{\sigma}(t),c^{\dagger}_{\sigma}(t')\right\}}$
and the lesser component
$G^{<}_{\sigma}(t,t')=i\braket{c^{\dagger}_{\sigma}(t')c_{\sigma}(t)}$.
In a thermal state these functions depend only on the difference~$t-t'=\tau$
and their Fourier transform is related by the fluctuation-dissipation theorem.
Out-of-equilibrium one can perform a Fourier transform with respect to~$\tau$
at fixed~$\bar t=(t+t')/2$~\cite{Eckstein2008} and obtain the spectral function
$A(\omega,\bar t)=-1/\pi\sum_\sigma \text{Im}G_\sigma^R(\omega,\bar t)$
and the occupation function
$N(\omega,\bar t)=i/(2\pi)\sum_\sigma G_\sigma^<(\omega,\bar t)$.
As a consequence of the time-dependent interaction, these functions have
oscillations in $\bar t$ with period $T=2\pi/\Omega$
and are even negative for some~$\omega$.
To extract meaningful information about the thermalization,
which happens on times~$\tau_\text{th}\gg T$, we average $A(\omega,\bar t)$
and $N(\omega,\bar t)$ over a few periods and obtain positive
$\bar A(\omega,\bar t)$ and $\bar N(\omega,\bar t)$
(cf. Supp. Mat.~\cite{Note1} Sec.~\ref{spec_func}).
The distribution function 
$\bar F(\omega,\bar t)=\bar N(\omega,\bar t)/\bar A(\omega,\bar t)$
provides a simple indicator for thermalization since in the thermal state 
it equals the Fermi-Dirac distribution.
In this case at early times we observe a non-thermal distribution
with a pseudo-periodic structure in~$\omega$ with period~$\Omega$.
This feature is related to the so-called Floquet subbands
characteristic of periodically driven systems~\cite{Tsuji2009}.
Then, at later times we observe a remarkably flat distribution
-- clearly the only one to be at the same time thermal
and pseudo-periodic.
This establishes that the fluctuation-dissipation relation is satisfied with
infinite temperature and therefore confirms thermalization.

\begin{figure*}
\includegraphics[width=\textwidth]{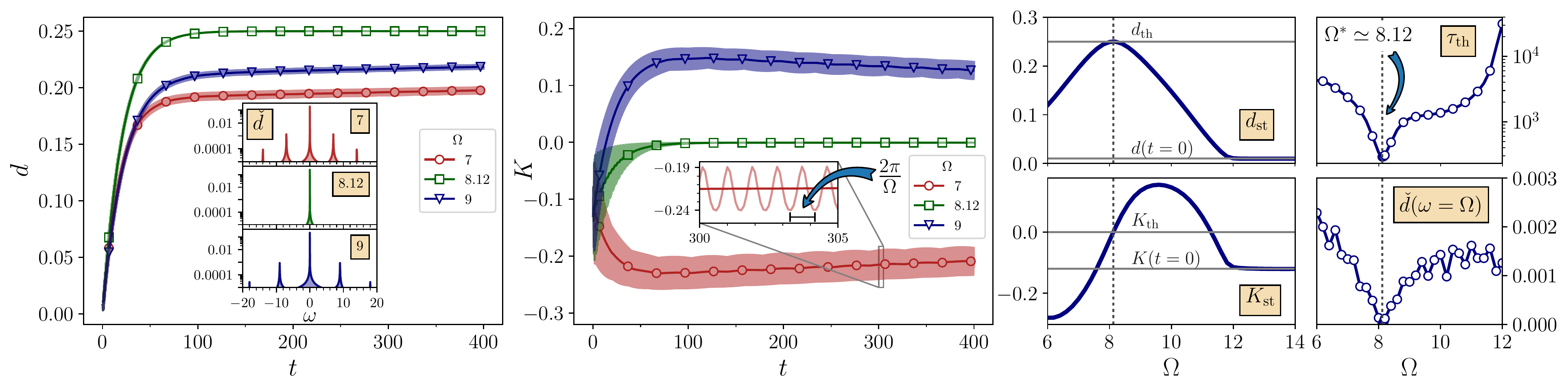}
\caption{\label{fig:fig2}
Floquet prethermalization and resonant thermalization ($U_0=8$).
Left panel:
Double occupancy~$d(t)$ for various drive frequencies~$\Omega$ shows
oscillations (shade) on top of an exponential relaxation (solid line).
Central panel: same for kinetic energy.
Inset of left panel: Fourier transform $\check d(\omega)$.
Right panels: Top left: stationary value $d_\text{st}(\Omega)$
with thermal value $d_\text{th}$ and initial value $d(t=0)$
for reference.
Bottom left: same for kinetic energy.
Top right: estimated thermalization time $\tau_\text{th}(\Omega)$.
Bottom right: weight of the peak $\check d(\omega=\Omega)$.
Dotted lines mark the resonant frequency $\Omega^*\simeq8.12$.
For $\Omega=7,9$ we see Floquet prethermalization with
$d_\text{st}\neq d_\text{th}$, $K_\text{st}\neq K_\text{th}$
and $\check d$ peaked at $\omega=\pm\Omega,\pm 2\Omega$.
For $\Omega=\Omega^*$ not only $d_\text{st}=d_\text{th}$
and $K_\text{st}=K_\text{th}$ but also the sharp minimum of
$\tau_\text{th}(\Omega)$ and the vanishing of $\check d(\omega=\Omega)$
signal the resonant thermalization.}
\end{figure*}

We now turn to the strong coupling regime at large average interaction
$U_0=8$ (Fig.~\ref{fig:fig2}).
The transition from moderate interaction appears
to be rather smooth (cf. Supp. Mat.~\cite{Note1} Sec.~\ref{amplitude}),
however for large interaction, in contrast with above, we find
qualitative differences as a function of the drive frequency.
As a first indication, while
for short times also in this case local observables oscillate on top of an
exponential relaxation,
now the stationary value depends on
frequency.
As we detail in the following, this signals the existence of different
dynamical regimes.
In particular, we find thermalization and damping of the oscillations
only for a critical frequency, which we estimate to be~$\Omega^*\simeq8.12$,
while for the other frequencies we observe a long-lived prethermal state.

For frequency below $\Omega^*$ the double occupancy and the kinetic
energy oscillate around an average which 
relaxes exponentially to a 
non-thermal plateau after the initial transient.
While for moderate interaction these oscillations damp out,
here they persist with constant amplitude. 
We calculate the Fourier transform
$\check d(\omega)=\int_{\tau_\text{pth}}^{t_\text{max}}
\diff{t}e^{i\omega t}d(t)$
where~$\tau_\text{pth}$ is the prethermalization time when the
plateau is attained and~$t_\text{max}$ is the maximum simulation time.
The peaks of~$\check d$ at multiples of~$\Omega$ demonstrate the
synchronization of the oscillations with the drive.
This, together with the non-thermal value of the plateau,
are the distinctive features of a
Floquet prethermal state in which the system appears to be trapped
for times longer than numerically accessible.
Since the plateau has a slight linear positive slope,
we can extrapolate it to intercept the thermal value~$d_\text{th}=0.25$
and in this way estimate a thermalization time~$\tau_\text{th}$
which turns out to be orders of magnitude
larger than at moderate interaction.

For frequency above $\Omega^*$ we find a very similar prethermalization
regime until, for $\Omega\simeq U_0+W=12$, we observe a 
sharp threshold behavior. This value corresponds to the maximum energy for
single-particle excitations above which
the system appears to be unable to absorb energy and local observables are
almost constant and equal to their initial equilibrium values.
Accordingly, 
the thermalization time grows exponentially with frequency,
in agreement with rigorous
bounds~\cite{Abanin2015,Abanin2017a}.
We remark also that, for a range of frequencies, the kinetic energy becomes
positive,
which is characteristic of a highly nonequilibrium state with population
inversion. While a similar phenomenon is observed in other Floquet
systems~\cite{Tsuji2011},  here it cannot be ascribed to an
effective change of sign of the interaction, since this would also cause
the double occupancy to increase above $0.25$.

The above picture radically changes for the critical frequency
$\Omega^*\simeq 8.12$ where
fast thermalization is found
despite the large interaction.
Here we observe an exponential relaxation of the double
occupancy and of the kinetic energy to the thermal values,
together with a full damping of oscillations.
At this specific frequency the Floquet prethermal state
is therefore melted away and the system is able to 
relax to the
infinite-temperature thermal state.
We name this phenomenon \emph{resonant} thermalization since for
$\Omega^*$ the periodic modulation of the interaction
is resonant with the energy~$\sim U_0$ of doublon excitations, i.e.
excitations that change the double occupancy.
This resonant condition allows 
the absorption of energy from the drive
and the creation of doublons,
which are otherwise suppressed by the large
average interaction
through a well-known bottleneck mechanism~\cite{Rosch2008a,Sensarma2010}.
Remarkably, the behavior of the system around $\Omega^*$ is strongly
reminiscent of a dynamical
transition~\cite{Eckstein2009b,Schiro2010,Tsuji2013a}.
This is clearly seen in the estimated thermalization
time~$\tau_\text{th}(\Omega)$
which has a sharp minimum for $\Omega^*$,
as well as from the peak at $\omega=\Omega$ of the Fourier
transform $\check d(\omega)$. The weight of this peak
goes to zero for $\Omega^*$ with singular behavior,
indicating the breakdown of synchronization and the approach
to the stationary thermal value.

\begin{figure}
\includegraphics[width=\columnwidth]{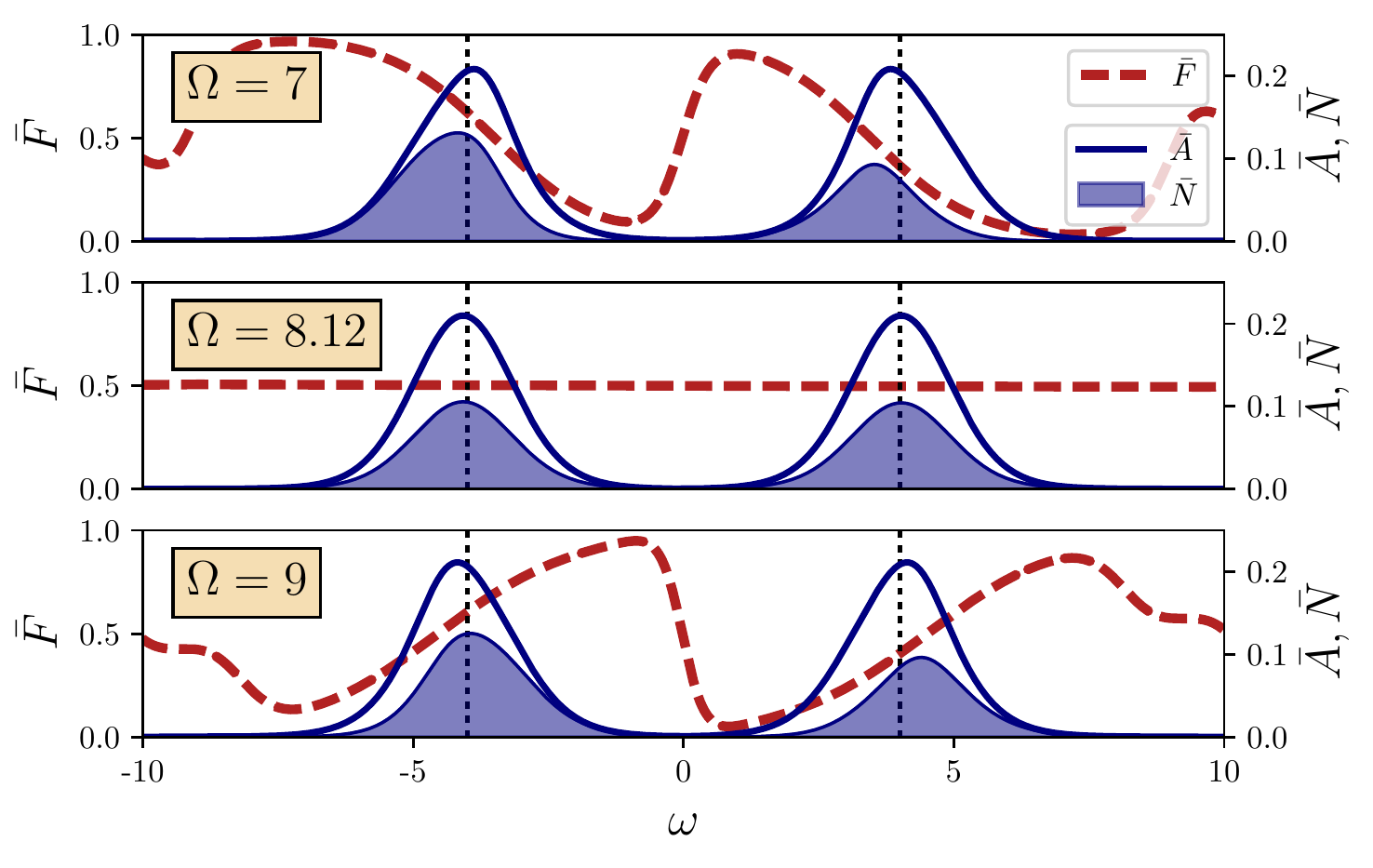}
\caption{\label{fig:fig3}
Averaged spectral function~$\bar A(\omega,\bar t)$,
occupation function~$\bar N(\omega,\bar t)$ and~$\bar F(\omega,\bar t)
=\bar N(\omega,\bar t)/\bar A(\omega,\bar t)$
for~$U_0=8$ and $\bar t=100$.
Prethermalization for $\Omega=7,9$
and thermalization for $\Omega=\Omega^*\simeq8.12$.
For $\Omega=9$ the population inversion is clear from the shift of $\bar N$
towards higher energy and the change of slop of $\bar F$ with respect
to $\Omega=7$. Dotted lines mark the approximate middle of the Hubbard bands.}
\end{figure}

The above results are corroborated by the evolution of the
spectral, occupation and distribution functions~(Fig.~\ref{fig:fig3}).
After the initial transient, these functions
reach a stationary state
independent of $\bar t$. This confirms that the plateau of the local
observables corresponds to a true steady state of the system.
For $\Omega\ne\Omega^*$ the distribution function $\bar F(\omega,\bar t)$
is clearly non-thermal and pseudo-$\Omega$-periodic,
as also found for the non-thermal transient at moderate interaction.
On the opposite, for the critical frequency $\Omega^*$ we find a
remarkably flat distribution which confirms
the thermalization at infinite temperature.
Interestingly, for $\Omega>\Omega^*$, corresponding to positive
kinetic energy, we indeed find a population inversion,
as it is clear from the shift towards high energy
of $\bar N$ and the change of slope of $\bar F$ with respect
to $\Omega<\Omega^*$ .

To gain an analytical insight into the Floquet prethermalization and the
resonant thermalization we use a
Floquet Schrieffer-Wolff
transformation~\cite{Schrieffer1966,Bukov2016,Wysokinski2017}.
This conveniently describes the strong coupling regime,
where doublon excitations are suppressed because of the large
average interaction, 
thus preventing the system from absorbing energy
unless the frequency of the drive is resonant with the doublon energy.
In practice, we introduce a time-periodic unitary
$R(t)=\exp S(t)$
which eliminates perturbatively in $V/U_0$ the terms
that do not conserve the double occupancy in
the transformed Hamiltonian
$\tilde H=e^{S}He^{-S}-i\partial_t S$.
This is obtained with an ansatz
$S(t)=(V/U_0)(\alpha(t) K_+-\alpha^*(t) K_-)$
where $\alpha(t)$ is a periodic function determined imposing the
vanishing of the commutator $[\tilde H,\sum_in_{i\up}n_{i\down}]$
up to terms of a given order in $V/U_0$,
and where we decompose the kinetic energy
in terms that do not change ($K_0$),
increase ($K_+$), or decrease ($K_-$) the double occupancy
(cf. Supp. Mat.~\cite{Note1} Sec.~\ref{s_w}).
For generic drive frequency the transformation is
well behaved and at first order in $V/U_0$ we find:
\begin{equation}
\label{d_sw}
\begin{split}
d(t)&= d(0)-2(V/U_0)\text{Re}[\alpha(T)\Tr(\rho(0)K_+)]\\
&+2(V/U_0)\text{Re}[
\alpha(t)e^{i\int_0^t U(t')dt'}\Tr(\rho(0)K_+(t))],
\end{split}
\end{equation}
where $K_+(t)\equiv e^{iVK_0t}K_+e^{-iVK_0t}$.
Eq.~\eqref{d_sw} captures the Floquet prethermal state
at long times multiples of $T=2\pi/\Omega$ (stroboscopic evolution)
when the double occupancy is synchronized with the drive
and oscillates around a frequency-dependent non-thermal value.
However, for the critical value $\Omega^*\simeq U_0$ and its submultiples,
the function $\alpha$ develops a singularity and the transformation breaks
down. This suggests that, at these frequencies, the Floquet prethermal state 
is unstable towards thermalization through non-perturbative processes 
 in $V/U_0$, as captured by DMFT.
Calculations at large interaction $U_0=14$ and drive
amplitude $\delta U=6$ clearly show
the resonant thermalization for frequencies $\Omega^*$ and
$\Omega^*/2$ (cf. Supp. Mat.~\cite{Note1} Sec.~\ref{resonant}).

The results we have presented here have a potential impact on various
experiments,
ranging from ultra-cold atoms in driven optical lattices, where one should
observe a sudden increase of the heating rate~\cite{Reitter2017}
at $\Omega=\Omega^*$;
to photo-excited organic Mott insulators~\cite{Singla2015}, where
one should observe a sudden filling of the gap in the
transient optical conductivity.
We also envisage further theoretical study, in particular on the effect of
non-local correlations in realistic lattices, which are likely to affect the
lifetime of the prethermal plateau.
Advances in the solution of the impurity problem would also be important,
as they would permit further investigations of the transition between moderate
and large interaction and the access to initial states at lower temperature.

In conclusion, to study periodically driven strongly correlated
electrons, we have considered the Fermi-Hubbard model
with time-periodic interaction.
Within nonequilibrium DMFT we have calculated the evolution of 
local observables and of the local Green function,
which provide evidence for thermalization or prethermalization.
We have showed the 
existence of three dynamical regimes:
(i)~Thermalization to infinite temperature at moderate interaction,
as expected for generic isolated quantum many-body systems;
(ii)~Floquet prethermalization at large
interaction, characterized by oscillations of local observables
around a non-thermal plateau
and a stationary non-thermal distribution function;
(iii)~Resonant thermalization at large interaction for an isolated critical
frequency~$\Omega^*$,
where local observables relax exponentially to the infinite-temperature thermal
value, together with a damping of oscillations and a flat distribution function.
We have then developed a periodic Schrieffer-Wolff transformation which
captures the qualitative features of the Floquet prethermal state
and whose breakdown for $\Omega^*$ indicates 
the non-perturbative nature of the resonant thermalization phenomenon.

\begin{acknowledgments}
This work is supported by the FP7/ERC, under Grant Agreement
No. 278472-MottMetals. MS acknowledges support from a  grant ``Investissements 
d’Avenir'' from  LabEx  PALM (ANR-10-LABX-0039-PALM) and from the CNRS
through the PICS-USA-14750.
\end{acknowledgments}

\bibliography{drivenFH}

\end{document}


\title{Supplemental Material for:\\
Resonant thermalization of periodically driven strongly correlated electrons}
\author{Francesco Peronaci}
\author{Marco Schir\'o}
\author{Olivier Parcollet}
\affiliation{Institut de Physique Th\'{e}orique (IPhT),
CEA, CNRS, UMR 3681, 91191 Gif-sur-Yvette, France}

\maketitle

The supplemental material is structured as follows.
In Sec.~\ref{amplitude} we present additional data and discuss the role
of the drive amplitude and average interaction.
In Sec.~\ref{resonant} we provide evidence for the thermalization at half
the resonant frequency $\Omega^*$. In Sec~\ref{nca_oca} we give the equations of the
non-crossing (NCA) and one-crossing (OCA) approximations and in
Sec.~\ref{benchmark} we provide OCA benchmark data at moderate interaction.
In Sec.~\ref{spec_func} we explain the analysis of the spectral function.
Finally, in Sec.~\ref{s_w} we discuss the Floquet Schrieffer-Wolff transformation.

\section{Role of drive amplitude and average interaction}
\label{amplitude}
In the main text we have considered a driven interaction
$U(t)=U_0+\delta U\sin\Omega t$ with drive amplitude $\delta U=2$ and 
average interaction $U_0=4$ and $U_0=8$.
Here we provide data which clarify the role of these parameters.
\begin{figure}[b]
\includegraphics[width=\columnwidth]{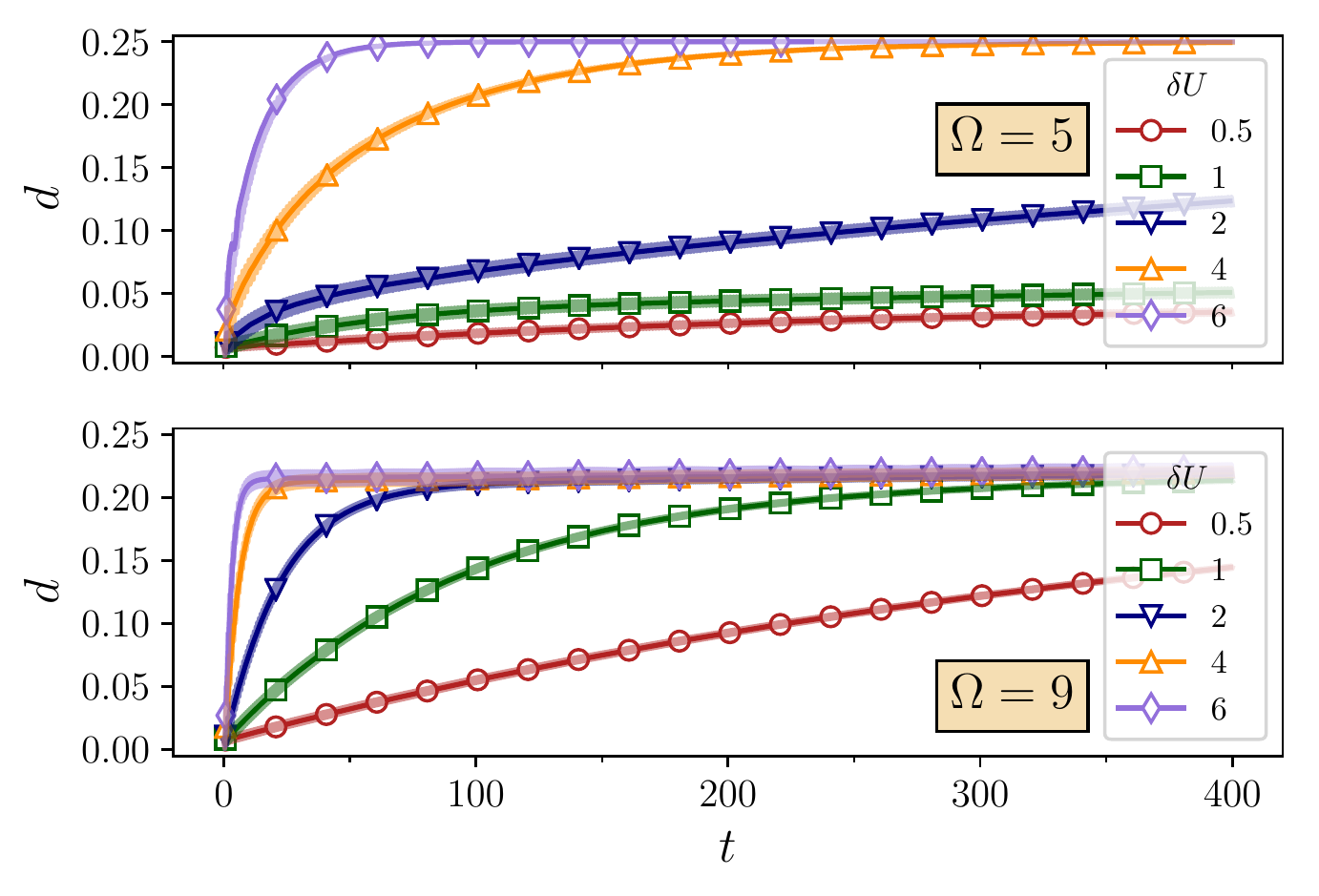}
\caption{\label{du}Role of drive amplitude at average interaction $U_0=8$.
Top panel: For drive frequency $\Omega=5$, well below the resonance $\Omega^*$,
a larger amplitude makes the system to thermalize.
Bottom panel: For frequency close to resonance ($\Omega=9$) a larger amplitude
gives a faster relaxation to the prethermal plateau, whose value does not
depend  on the drive amplitude.}
\end{figure}
In Figure~\ref{du} we plot the time evolution of the double occupancy
at fixed large interaction and different drive amplitudes.
Quite interestingly, a larger drive amplitude has different effects
for drive frequency far from resonance or close to resonance.
In the former case ($\Omega=5$) we observe that a large drive amplitude
makes the system thermalize despite the large average interaction.
As opposite, near resonance ($\Omega=9$)
we observe a faster dynamics which, however, eventually
saturates to the same prethermal plateau described in the main text.
Remarkably, the value of this plateau does not depend on the drive
amplitude, as opposed to its strong dependence on the drive frequency, as
discussed in the main text.

\begin{figure}[t]
\includegraphics[width=0.8\columnwidth]{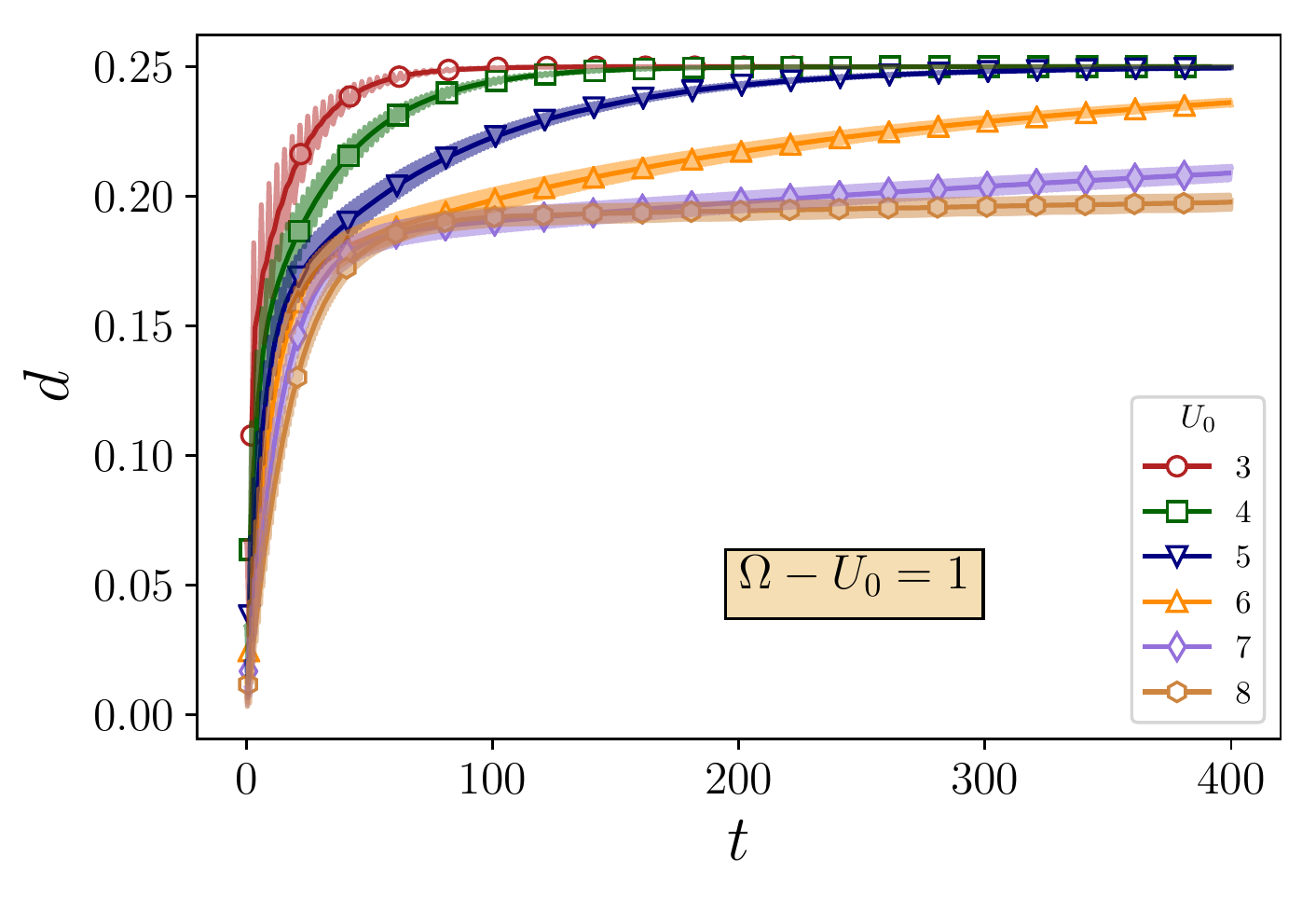}
\caption{Evolution from moderate to large interaction at fixed $\Omega-U_0=1$.
A kink is visible at $\tau_\text{pth}\simeq 25$ for $U_0=5,6$
at which the dynamics slows down.}
\label{u0}
\end{figure}

A second interesting question concerns the connection of the regimes at moderate
and large interactions discussed in the main text.
In Figure~\ref{u0} we plot the time evolution of the double occupancy
with different average interactions and at a fixed difference $U_0-\Omega=1$.
In this way we can disentangle the effect
of a different interaction from an effective decrease of the drive frequency.
As the interaction is increased from $U_0=3$ to $U_0=8$, we observe a rather
smooth change, with the appearance of a well visible kink at
$\tau_\text{pth}\simeq 25$ ($U_0=5,6$) at which the dynamics slows down,
and which eventually turns to a prethermal plateau.

\section{Resonant Thermalization at $\Omega^*/2$}
\label{resonant}
The Floquet Schrieffer-Wolff transformation used in the main text and further
discussed below, shows that the Floquet prethermal
state is unstable towards thermalization whenever the drive frequency
is resonant with the doublon excitation energy $\Omega^*$ \emph{and its submultiples}.
This suggests that around frequencies $\frac{\Omega^*}{n}$ with $n$ integer we should
observe a phenomenon similar to what discussed in the main text around $\Omega^*$.

To confirm this picture, we present data with large average interaction
$U_0=14$ and drive amplitude $\delta U=6$. With these values the two resonances
at $\Omega^*$ and $\frac{\Omega^*}{2}$ are well separated and, importantly, the frequency
$\frac{\Omega^*}{2}$ is still reasonably high such that we can numerically
access the time scale at which a prethermal plateau is well visible.
In Figure~\ref{du} we provide evidence for the resonant thermalization
at $\Omega^*\simeq13.6$ and $\frac{\Omega^*}{2}\simeq6.8$.
Similarly to what discussed in the main text for $U_0=8$, the double occupancy
relaxes to a prethermal plateau whose value is non-monotonic in the drive frequency.
Around both $\Omega^*$ and $\frac{\Omega^*}{2}$ however, the double occupancy relaxes
to the thermal value and, importantly, the amplitude of its oscillations vanishes.
This second resonance is not observed at $U_0=8$ because in this case at
$\frac{\Omega^*}{2}=4$ the dynamics becomes much slower and a clear signature
of prethermalization cannot be obtained
within the numerically accessible time scales.
\begin{figure}
\includegraphics[width=\columnwidth]{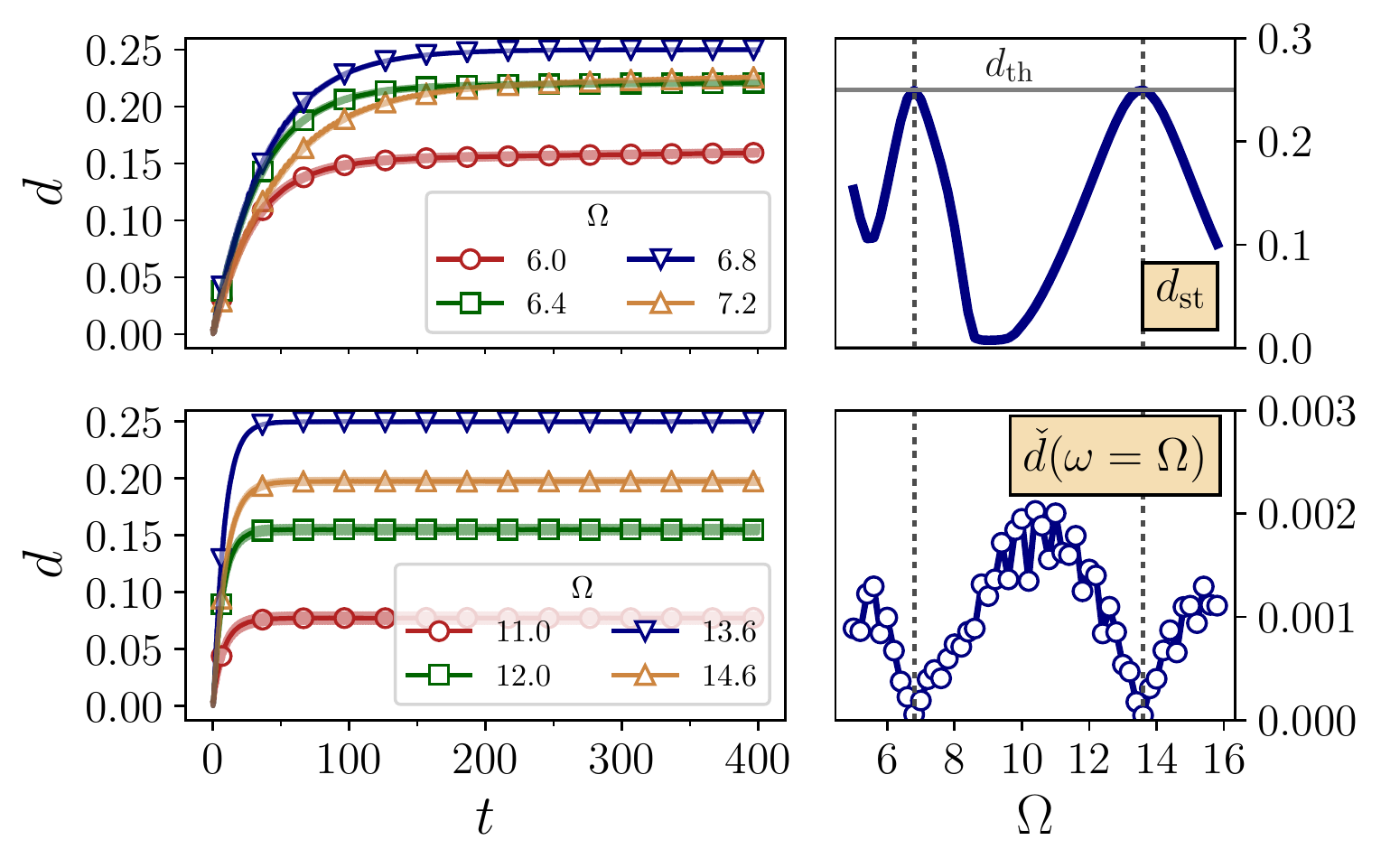}
\caption{Floquet prethermalization and resonant thermalization at $U_0=14$ and $\delta U=6$.
Left panels: double occupancy $d(t)$ shows a long-lived prethermal plateau eventually
evolving to the thermal value as the frequency is swept across 
$\frac{\Omega^*}{2}\simeq6.8$ (top) and $\Omega^*\simeq13.6$ (bottom).
Top right panel: stationary value $d_\text{st}(\Omega)$.
Bottom right panel: weight of the Fourier component $\check d(\omega=\Omega)$ at the
frequency of the drive.}
\end{figure}

\section{Non-crossing and one-crossing approximations}
\label{nca_oca}
The non-crossing approximation (NCA) and one-crossing approximations (OCA)
are based on the introduction of the atomic-state propagator $R$
through the following Dyson equation:
\begin{subequations}
\begin{gather}
\label{dyson_eq}
i\partial_z R(z,z')=H_\text{loc}(z)R(z,z')+(S\otimes R)(z,z'),\\
R(z,z^-)=-i\I,
\end{gather}
\end{subequations}
where $H_\text{loc}(z)$
is the local Hamiltonian
and $z,z'$ are variables on the three branch Keldysh contour $\mathcal{C}$,
with $z^-$ infinitesimally before $z$ on $\mathcal{C}$.

In Eq.~\eqref{dyson_eq} $S$ is the self-energy of the atomic-state propagator,
which describes the effects of the nonequilibrium bath on the dynamics of
the impurity and can be expanded in powers of the hybridization~$\Delta_\s$.
The NCA consists in a self-consistent truncation of this series at the first order:
\begin{equation}
\begin{split}
\label{nca_self}
S^{(1)}(z,z')&=i\sum_{\sigma=\up,\down}c_\s\dagg R(z,z')c_\s \Delta_\s(z,z')\\
&-i\sum_{\sigma=\up,\down}c_\s R(z,z')c_\s\dagg\Delta_\s(z',z),
\end{split}
\end{equation}
while within the OCA one adds also the self-consistent second-order contributions:
\begin{equation}
\begin{split}
\label{oca_self}
&S^{(2)}(z,z')=
\sum_{\sigma\sigma'}\int_{z'}^z\diff{z_1}\int_{z'}^{z_1}\diff{z_2}\\
\Bigl[&
c_\s\dagg R(z,z_1)c_{\s'}\dagg R(z_1,z_2)c_\s R(z_2,z')c_{\s'}
\Delta_\s(z,z_2)\Delta_{\s'}(z_1,z')\\
-&c_\s\dagg R(z,z_1)c_{\s'} R(z_1,z_2)c_\s R(z_2,z')c_{\s'}\dagg
\Delta_\s(z,z_2)\Delta_{\s'}(z',z_1)\\
-&c_\s R(z,z_1)c_{\s'}\dagg R(z_1,z_2)c_\s\dagg R(z_2,z')c_{\s'}
\Delta_\s(z_2,z)\Delta_{\s'}(z_1,z')\\
+&c_\s R(z,z_1)c_{\s'} R(z_1,z_2)c_\s\dagg R(z_2,z')c_{\s'}\dagg
\Delta_\s(z_2,z)\Delta_{\s'}(z',z_1)
\Bigr].
\end{split}
\end{equation}

In Eq.~\eqref{dyson_eq} it appears the following \emph{clockwise} contour convolution:
\begin{equation}
\label{conv}
(A\otimes B)(z,z')=
\begin{cases}
\int_{z'}^z\diff{\bar z}A(z,\bar z)B(\bar z,z') & \text{if $z>z'$},\\
\int_{0^+}^z\diff{\bar z}A(z,\bar z)B(\bar z,z')\\
\,+\int^{-i\beta}_{z'}\diff{\bar z}A(z,\bar z)B(\bar z,z')
& \text{if $z<z'$},
\end{cases}
\end{equation}
where $0^+$ and $-i\beta$ are, respectively, the first and last point
of~$\mathcal{C}$. Moreover it should be stressed that
in Eqs.~\eqref{dyson_eq} and \eqref{nca_self} the products
among~$R$, $H_\text{loc}$, $S$, $c_\s$
and $c_\s\dagg$ are to be understood as matrix products in the
impurity Hilbert space.
Since $S$ depends on $R$ in practice one has to solve iteratively
Eqs.~\eqref{dyson_eq} and \eqref{nca_self} until a self-consistent solution
is found.
Moreover, this guarantees that the NCA and OCA are conserving approximations.

From the atomic-state propagator $R$ we can calculate all the thermodynamics
quantities of the impurity:
\begin{gather}
\label{part_func}
Z=\Tr(iR(-i\beta,0^+)),\\
\label{density}
n_\s(z)=\Tr(\xi iR(z^-,z)c_\s\dagg c_\s)Z^{-1},\\
d(z)=\Tr(\xi iR(z^-,z)c_\up\dagg c_\up c_\dn\dagg c_\dn)Z^{-1},
\end{gather}
where the trace is over the impurity Hilbert space, 
and $\xi_{mn}=\braket{m|\xi|n}=\pm\delta_{mn}$ if $\ket{m}$
contains an even or odd number of fermions.
The equation for the Green function depends on the approximation used and
within the NCA reads:
\begin{equation}
\label{green_function_nca}
G_\s^{(1)}(z,z')=i\Tr(\xi R(z',z)c_\s R(z,z')c_\s\dagg)Z^{-1},
\end{equation}
while within the OCA one adds also the terms: 
\begin{equation}
\begin{split}
\label{green_function_nca}
&G_\s^{(2)}(z,z')=Z\inv\sum_{\s'}\int_z^{z'}\diff{z_1}\int_{z'}^{z}\diff{z_2}\\
\Bigl[&
\Delta(z_1,z_2)\Tr(\xi R(z',z_1)c_{\s'}\dagg R(z_1,z)c_\s R(z,z_2)c_{\s'}
R(z_2,z')c_\s\dagg)\\
-&\Delta(z_2,z_1)\Tr(\xi R(z',z_1)c_{\s'} R(z_1,z)c_\s R(z,z_2)c_{\s'}\dagg
R(z_2,z')c_\s\dagg\Bigr].
\end{split}
\end{equation}

The first step of the implementation is the projection of the various equations
onto a basis of the impurity Hilbert space.
In this case we can conveniently use the
basis $\{\ket{m}\}=\{\ket{0},\,\ket{\up},\,\ket{\dn},\,\ket{\up\dn}\}$
on which $R$, $H_\text{loc}$ and $S$ are diagonal.
The second step is the projection of the equations
onto the different Keldysh components, which can be made by means
of the following
Langreth rules for the clockwise convolution Eq.~\eqref{conv}:
\begin{subequations}
\label{langreth}
\begin{align}
\label{langreth1}
(A\otimes B)\gtr(t,t')&=
\int_{t'}^t\diff{\bar t}A^>(t,\bar t)B^>(\bar t,t'),\\
\label{langreth2}
\nonumber
(A\otimes B)\mix(t,\tau)&=
\int_{0}^t\diff{\bar t}A\gtr(t,\bar t)B\mix(\bar t,\tau)\\&+
\int_\tau^\beta\diff{\bar\tau}A\mix(t,\bar\tau)B\mat(\bar\tau-\tau),\\
\label{langreth3}
\nonumber
(A\otimes B)\les(t,t')&=
\int_{0}^t\diff{\bar t}A\gtr(t,\bar t)B\les(\bar t,t')\\
\nonumber
&-\int_{0}^{t'}\diff{\bar t}A\les(t,\bar t)B^>(\bar t,t')
\\&-i\int_0^\beta\diff{\bar\tau}A\mix(t,\bar\tau)B\miy(\bar\tau,t'),\\
\label{langreth4}
(A\otimes B)\mat(\tau)&=
\int_{0}^\tau\diff{\bar \tau}A\mat(\tau-\bar\tau)B\mat(\bar\tau),
\end{align}
\end{subequations}
which differ with respect to the rules for the usual convolution in
two aspects: (i) the kernel in Eqs.~\eqref{langreth1} to
\eqref{langreth3} is the greater component instead of the retarded;
(ii) the $\tau$-integrals in Eqs.~\eqref{langreth2} and \eqref{langreth4}
are over a limited imaginary-time interval.

Importantly, under hermitian conjugation $R$
satisfies the same properties of a bosonic or fermionic Green function:
\begin{align}
(R^\gtrless(t,t'))\dagg &=-R^\gtrless(t',t),\\
(R\mix(t,\tau))\dagg &=-\xi R\miy(\beta-\tau,t),
\end{align}
and this can be used to restrict the calculation to half of the
$(t,t')$ plane and to get $R\miy$ from $R\mix$.

Finally, one needs the initial condition for the atomic-state propagator,
which is obtained from 
$R\mat$ -- whose equation is uncoupled from the other Keldysh components --
through the following relations:
\begin{subequations}
\begin{align}
R\gtr(0,0)&=-i\I,\\
R\les(0,0)&=i\xi R\mat(\beta),\\
R\mix(0,\tau)&=i\xi R\mat(\beta-\tau).
\end{align}
\end{subequations}

\section{OCA benchmark at moderate interaction}
\label{benchmark}
\begin{figure}
\includegraphics[width=0.8\columnwidth]{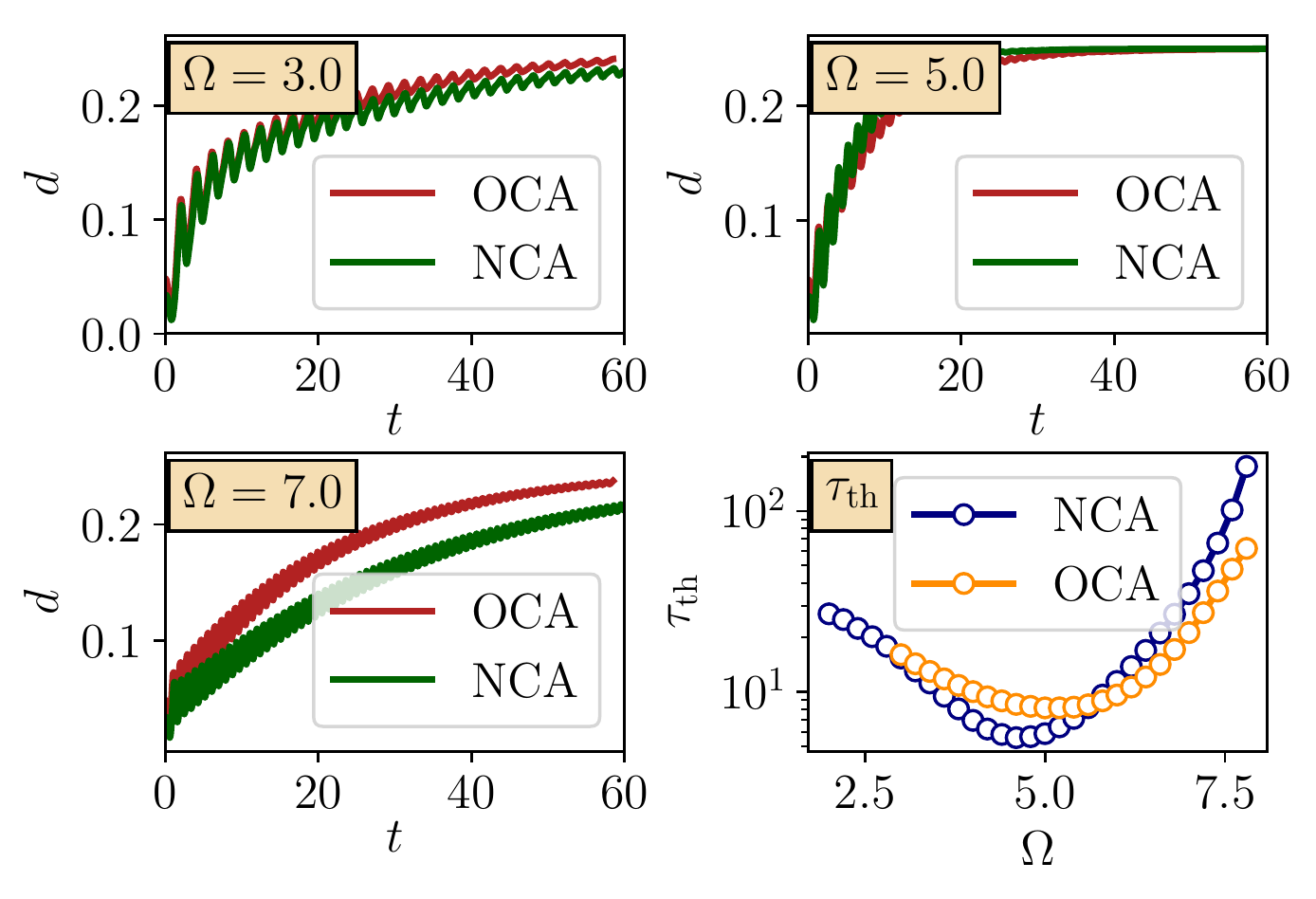}
\caption{Comparison NCA-OCA at moderate interaction $U_0=4$
and drive amplitude $\delta U=2$.
Top left, top right and bottom left panels: double occupancy $d(t)$ for different
drive frequency and within OCA and NCA.
Bottom right panel: thermalization time $\tau_\text{th}(\Omega)$ for OCA and NCA.}
\label{fig_sm6}
\end{figure}

We have performed benchmark calculations at moderate interaction $U_0=4$.
In Figure~\ref{fig_sm6} we plot the time evolution of the double occupancy
for three different frequencies $\Omega=3,5,7$ showing that
the OCA corrections are only quantitative and, importantly, also in this case
we observe thermalization.
In the same figure, we compare the thermalization times within NCA and OCA,
as extracted with an exponential fit of the time evolution.
The OCA results thus shows the same qualitative features of the NCA,
and slight quantitative differences such as a shift of 
the frequency at which the thermalization time $\tau_\text{th}$
is minimum from $\simeq4.8$ to $\simeq5.5$.

\section{Spectral function analysis}
\label{spec_func}
\begin{figure}
\includegraphics[width=0.8\columnwidth]{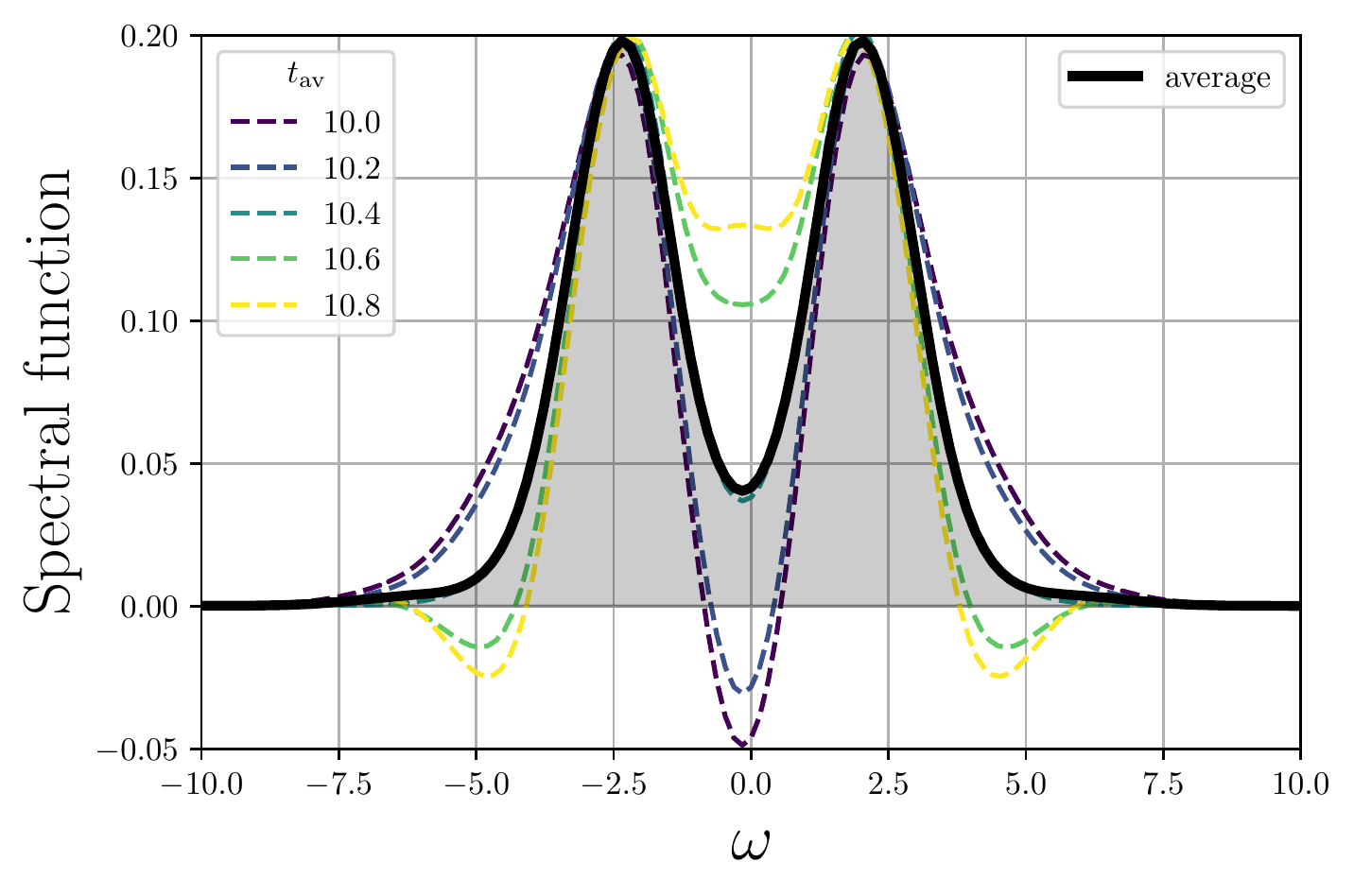}
\caption{
Spectral function $\bar A(\omega,\bar t)$ for $U_0=4$, $\delta U=2$
and $\Omega=4$ for different
$\bar t$ (dashed lines) and average over $n=5$ periods
(solid line).}
\label{fig_sm}
\end{figure}

Let us consider the retarded component of the Green function:
\begin{equation}
G^{R}_{\sigma}(t,t')=-i\theta(t-t')
\braket{\left\{c_{\sigma}(t),c^{\dagger}_{\sigma}(t')\right\}},
\end{equation}
from which we calculate the time evolution of the spectral function 
discussed in the main text.
The same analysis was carried out for the lesser component to derive
the time evolution of the occupation function.
In a thermal state $G^{R}_{\sigma}(t,t')$ depends only on the
difference~$t-t'=\tau$.
Out-of-equilibrium one can perform a Fourier transform with respect to~$\tau$
at fixed~$\bar t=(t+t')/2$:
\begin{equation}
A(\omega,\bar t)=-\frac{1}{\pi}\sum_\sigma \text{Im}
\int\diff{\tau}e^{i\omega\tau}G_\sigma^R(\bar t+\frac{\tau}{2},
\bar t-\frac{\tau}{2}).
\end{equation}
As a consequence of the time-dependent interaction, this function has
oscillations in $\bar t$ with period $T=2\pi/\Omega$
and is even negative for some~$\omega$.
However we can average it over a few periods:
$\bar A(\omega,\bar t)=1/(nT)\int_{\bar t}^{\bar t+nT}A(\omega,\bar t')
\diff{\bar t'}$
and obtain a positive spectral function (Fig.~\ref{fig_sm}).

\section{Floquet Schrieffer-Wolff transformation}
\label{s_w}
To derive the Floquet Schrieffer-Wolff transformation it is convenient
to rewrite the Hamiltonian Eq. (1) of the main text as:
\begin{equation}
\label{ham}
H(t)=U(t)D-V(K_0+K_++K_-),
\end{equation}
with the definition of the following operators: 
\begin{subequations}
\label{kinetic}
\begin{gather}
D=\sum_i(n_{i\up}-\frac{1}{2})(n_{i\dn}-\frac{1}{2}),\\
K_0=\sum_{\braket{ij}\s}c_{i\s}\dagg c_{j\s}(n_{i\bar\s}n_{j\bar\s}+
(1-n_{i\bar\s})(1-n_{j\bar\s})),\\
K_+=(K_-)\dagg=\sum_{\braket{ij}\s}c_{i\s}\dagg c_{j\s}n_{i\bar\s}(1-n_{j\bar\s}),
\end{gather}
\end{subequations}
where with $\braket{ij}$ we indicate nearest-neighbor sites for
which~$V_{ij}=-V$.
The double occupancy then reads $d(t)=\Tr\rho(t)D$
and we have decomposed the kinetic energy
in terms that do not change the double occupancy ($K_0$),
increase it ($K_+$) or decrease it ($K_-$), as it is clear
from the following commutators:
\begin{equation}
\label{comm_rules}
[D,K_0]=0,\quad\quad
[D,K_\pm]=\pm K_\pm.
\end{equation}
\begin{figure}
\includegraphics[width=0.7\columnwidth]{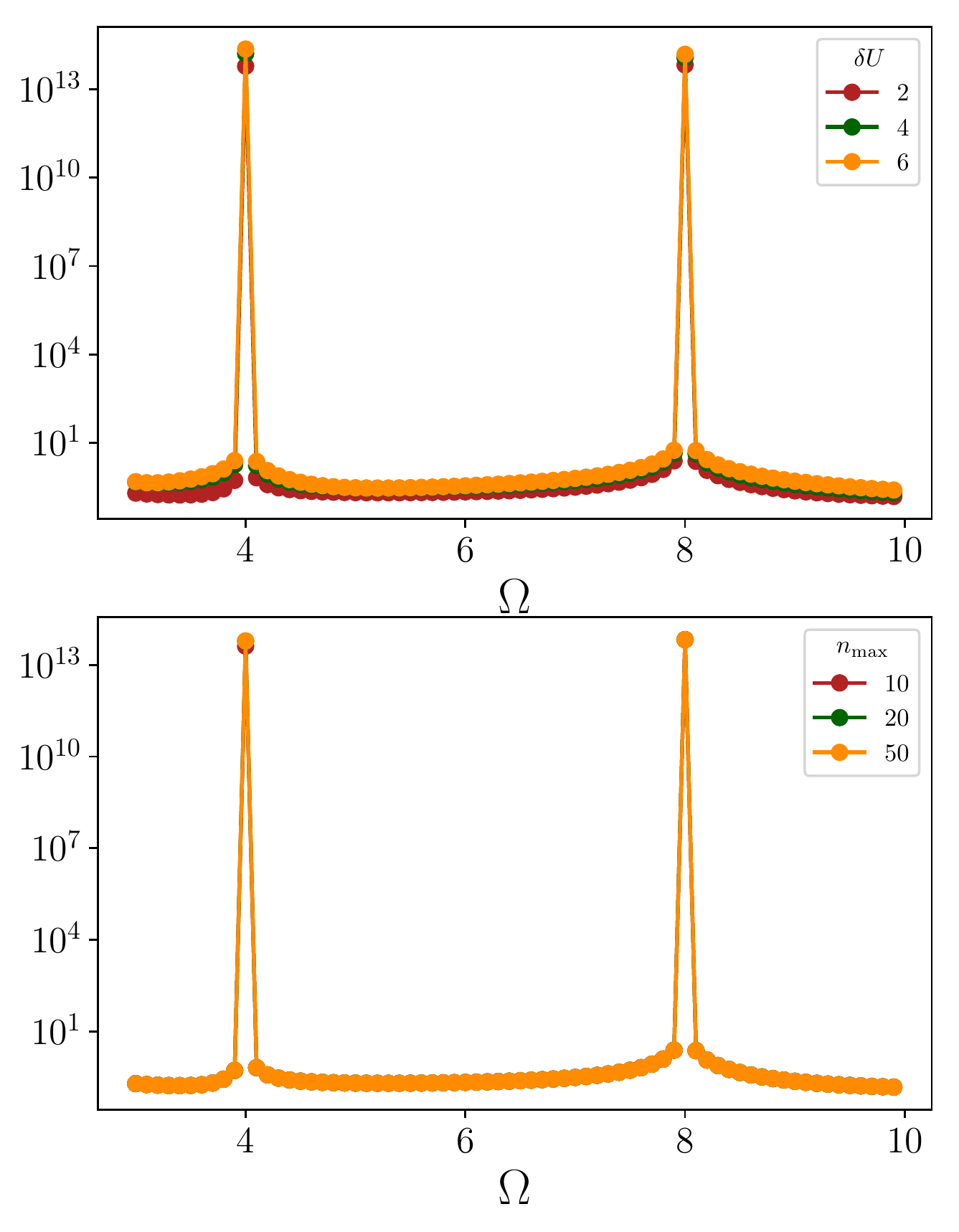}
\caption{Break down of the Schrieffer-Wolff transform. Plot of
$|\sum_n\alpha_n|$ as a function of the frequency of the drive $\Omega$
for $U_0=8$ and for different $\delta U$ (top panel, fixed $n_\text{max}=50$)
or different $n_\text{max}$ (bottom panel, fixed $\delta U=2$).}
\label{sf}
\end{figure}
In the limit $V/U_0=0$ the double occupancy is a conserved quantity.
To calculate the corrections in the case of small but finite $V/U_0$
we perform a change of basis:
\begin{equation}
\begin{split}
\bar\rho(t)
&\equiv R(t)\rho(t)R\dagg(t)\\
&=R(t)U(t)\rho(0) U\dagg(t)R\dagg(t)\\
&=R(t)U(t)R\dagg(0)\bar\rho(0) R(0)U\dagg(t)R\dagg(t)\\
&=\bar U(t)\bar\rho(0) \bar U\dagg(t),
\end{split}
\end{equation}
where $U(t)$ and $\bar U(t)=R(t)U(t)R\dagg(0)$ are the evolution operators
in the original and in the new basis and:
\begin{gather}
i\partial_t \bar U_t=\bar H(t)\bar U(t)\\
\label{tilde_ham}
\begin{split}
\tilde H(t)&=R(t) H(t)R\dagg(t)-iR(t)\partial_t R\dagg(t)\\
&=e^{-S(t)}H(t)e^{S(t)}-i\partial_t S(t)
\end{split}
\end{gather}
where we introduce the representation
$R(t)=e^{-S(t)}$ with $S\dagg(t)=-S(t)$.
We can choose $S(t)$ in such a way to eliminate perturbatively
in $V/U_0$ the terms in $\tilde H(t)$ that do not commute with $D$:
\begin{equation}
\begin{split}
\tilde H(t)&=H(t)+[H(t),S(t)]-i\partial_t S(t)+\mathcal{O}(V^2/U_0)\\
&=U(t)D-V(K_0+K_++K_-)\\
&+U(t)[D,S(t)]-i\partial_t S(t)+\mathcal{O}(V^2/U_0).
\end{split}
\end{equation}
Therefore at first order we find the condition:
\begin{equation}
\label{condition}
-V(K_++K_-)+U(t)[D,S(t)]-i\partial_t S(t)=0.
\end{equation}

If we now make a periodic ansatz of the following form
inspired by the standard Schrieffer-Wolff transformation:
\begin{equation}
S(t)
=\frac{V}{U_0}\sum_n e^{-in\Omega t}(\alpha_nK_+-\alpha^*_{-n}K_-),
\end{equation}
then the condition Eq.~\eqref{condition} becomes:
\begin{equation}
-U_0\delta_{n,0}+\sum_mU_{n-m}\alpha_m-n\Omega\alpha_n=0,
\end{equation}
where $U_n$ is the Fourier component of $U(t)$
and in particular if $U(t)=U_0+\delta U\sin(\Omega t)$
we obtain:
\begin{equation}
\label{system}
(n\Omega-U_0)\alpha_n+i\frac{\delta U}{2}(\alpha_{n+1}-\alpha_{n-1})
=U_0\delta_{n,0}.
\end{equation}
If the system Eq.~\eqref{system} is not singular,
we can finally calculate the
correction of order $V/U_0$ to the double occupancy:
\begin{equation}
\begin{split}
d(t)
&=\Tr[R(0)\rho(0)R\dagg(0)\bar U\dagg(t)R(t)DR\dagg(t)\bar U(t)]\\
&=\Tr(\bar \rho(0)\bar U\dagg(t)D\bar U(t))\\
&+\Tr([\rho(0),S(0)]\bar U\dagg(t)D\bar U(t))\\
&+\Tr(\rho(0)\bar U\dagg(t)[D,S(t)]\bar U(t)).
\end{split}
\end{equation}
which coincides term by term with Eq.~(3) of the main text.

On the other hand, if the matrix $A_{nm}$ is singular the above calculation
breaks down. We have solved numerically this problem with a cut-off to the
matrix size and we show in Fig.~\ref{sf} that this break down happens for a
frequency of the drive $\Omega\simeq \frac{U_0}{n}$ with
integer $n$.
For this frequencies, despite the small $V/U_0$,
the corrections to a constant double occupancy are non-perturbative in $V/U_0$
and this explains why the system does not reach a prethermal
plateau but instead is able to thermalize.